\documentclass[aps,prd,onecolumn,nofootinbib]{revtex4}
\usepackage{graphicx}% Include figure files
\usepackage{bm}% bold math
\usepackage{amsfonts}
\usepackage{latexsym}
\usepackage{amsmath}
\usepackage[english]{babel}
\usepackage[latin1]{inputenc}
\usepackage[T1]{fontenc}
\usepackage{slashed}
\usepackage{float}
\usepackage{subfigure}
\usepackage{color}

  \newcommand{\be}{\begin{equation}}
  \newcommand{\ee}{\end{equation}}
  \newcommand{\ba}{\begin{eqnarray}}
  \newcommand{\ea}{\end{eqnarray}}
  \newcommand{\eite}{\end{itemize}}
  \newcommand{\bite}{\begin{itemize}}

\def\ltap{\ \raisebox{-.4ex}{\rlap{$\sim$}} \raisebox{.4ex}{$<$}\ }

 \newcommand{\meff}{\mbox{$\left|  < \!  m \!  > \right| \ $}}

\definecolor{orange}{rgb}{1,0.5,0}

\begin{document}

% \preprint{SISSA 02/2012/EP,~CFTP/12-001}
\rightline{SISSA 02/2012/EP,~CFTP/12-001}

\title{Revisiting Leptogenesis 
in a SUSY $SU(5) \times T^{\prime}$ Model of Flavour}

\author{A. Meroni}
\email[]{ameroni@sissa.it}
\affiliation{SISSA and INFN-sezione di Trieste, 
Via Bonomea 265, 34136 Trieste, Italy}
% \affiliation{Instituto Nazionale di Fisica Nucleare, 
% Sezione di Trieste, Via Valerio 2, 34126 Trieste, Italy}

\author{E. Molinaro}
\email[]{emiliano.molinaro@ist.utl.pt}
\affiliation{Centro de F\'isica Te\'orica de Part\'iculas (CFTP), 
Instituto Superior T\'ecnico,\\
Technical University of Lisbon, 1049-001 Lisboa, Portugal}

\author{S. T. Petcov
\footnote{Also at: Institute of Nuclear Research and
Nuclear Energy, Bulgarian Academy of Sciences, 
1784 Sofia, Bulgaria}
}
% \email[]{petcov@sissa.it}
\affiliation{SISSA and INFN-sezione di Trieste, Via Bonomea 265, 
34136 Trieste, Italy}
% \affiliation{Instituto Nazionale di Fisica Nucleare, Sezione 
% di Trieste, Via Valerio 2, 34126 Trieste, Italy}
\affiliation{IPMU, University of Tokyo, Tokyo, Japan}

% \date{today}

\begin{abstract}
We investigate the generation of the 
baryon asymmetry of the Universe
within a SUSY $SU(5) \times T^{\prime}$ model of flavour, 
which gives rise to realistic masses and mixing 
patterns for quarks and leptons.
The model employs the see-saw mechanism 
for generation of the light neutrino masses 
and the baryon asymmetry is produced via 
leptogenesis. We perform detailed calculations of both 
the CP violating lepton asymmetries, originating 
from the decays of the heavy Majorana neutrinos
operative in the see-saw mechanism,
and of the efficiency factors 
which account for the lepton asymmetry wash-out 
processes in the Early Universe.
The latter are calculated by solving numerically 
the system of Boltzmann equations describing the 
generation and the evolution of the lepton asymmetries.
The baryon asymmetry in the model considered 
is proportional to the $J_{CP}$ factor, 
which determines the magnitude of CP violation 
effects in the oscillations of flavour neutrinos.
The leptogenesis scale can be sufficiently 
low, allowing to avoid the potential gravitino problem.
\end{abstract}

\maketitle

%%%%%%%%%%%%%%%%%%%%%%%%%%%%%%
%
\section{Introduction}
%
%%%%%%%%%%%%%%%%%%%%%%%%%%%%
%
 In the present article we consider the generation 
of the baryon asymmetry of the Universe in the 
SUSY model of flavour based on the 
$SU(5)\times T'$ symmetry, which was developed
in \cite{Chen:2007afa,Chen:2009gf}.
The model possesses a number of appealing features 
which makes it worthwhile to investigate 
whether it provides also a viable 
scenario for the baryon asymmetry generation. 

 The group $T'$ is the double covering of the 
symmetry group of the tetrahedron $A_4$ 
(see, e.g., \cite{Ishimori:2010au}).
It was realised by a number of authors
(see, e.g., \cite{frampton}) 
that the $T'$ symmetry can be used for 
the description of masses and mixing 
of both leptons and quarks.
The  $SU(5)\times T'$ model of flavour 
of interest accounts successfully  
for the pattern of quark masses and mixing, 
including the CP violation in the quark sector 
\cite{Chen:2009gf}.
It is free of discrete gauge anomalies
~\cite{Luhn:2008xh}
and gives rise to realistic masses and 
mixing of the leptons as well.

  The $SU(5)\times T'$ model 
proposed in \cite{Chen:2007afa,Chen:2009gf} 
we will discuss in the present article, 
includes three right-handed (RH) neutrino 
fields $N_{lR}$, $l=e,\mu,\tau$, 
which possess a Majorana mass term. 
The light neutrino masses 
are generated by the type I see-saw 
mechanism and are naturally small.
The light neutrino mass spectrum is 
predicted \cite{AMSP1109} to be with 
normal ordering and is 
hierarchical (throughout this article we use 
the definitions and the conventions 
given in \cite{PDG10}).
The neutrino masses $m_j$, $j=1,2,3$,
are functions of two real parameters of the 
model \cite{Chen:2009gf,AMSP1109}. 
The latter can be determined 
by using the values of the two neutrino mass 
squared differences, $\Delta m^2_{21}$ and  
$\Delta m^2_{31}$, or of  $\Delta m^2_{21}$ and 
the ratio $r = \Delta m^2_{21}/\Delta m^2_{31}$,
obtained in the global analyses of the 
neutrino oscillation data.
Using the best fit values of 
$\Delta m^2_{21} = 7.58\times 10^{-5}~{\rm eV^2}$ and  
$\Delta m^2_{31} = 2.35\times 10^{-3}~{\rm eV^2}$,
found in the analysis performed in \cite{Fogli:2011qn}, 
we have \cite{AMSP1109}: 
%%%%%%%%%%%%%%%%%%%%%%%%%%
\be 
m_1=1.14 \times 10^{-3}\,{\rm eV}, \quad m_2=8.78 \times 10^{-3}\,{\rm eV},
\quad m_3=4.867 \times 10^{-2}\,{\rm eV}. 
\label{mj}
\ee
%%%%%%%%%%%%%%%%%%%%%%%%%%
%
The values of  $m_1$, $m_2$ and $m_3$
are essentially fixed:
the uncertainties corresponding to the 
3$\sigma$ ranges of allowed values of 
$\Delta m^2_{21}$ and $r$
are remarkably small
\cite{AMSP1109}.

 The part of the Pontecorvo, Maki, Nakgawa and 
Sakata (PMNS) neutrino mixing matrix (see \cite{PDG10}), 
resulting from the diagonalisation of 
the Majorana mass term of the 
left-handed flavour neutrino fields $\nu_{lL}(x)$, 
$l=e,\mu,\tau$, which is generated by the 
see-saw mechanism, is of the tri-bimaximal 
form~\cite{tri1}. 
The latter is ``corrected'' by the unitary matrix
originating from the diagonalisation of the 
charged lepton mass matrix $M_{e}$ 
(for a general discussion of such corrections
see, e.g., \cite{FPR,HPR07,Marzocca:2011dh}). 
Since the model is based on the $SU(5)$ GUT 
symmetry, the charged lepton mass matrix 
is related to the down-quark 
mass matrix $M_d$. The model exploits the 
Georgi-Jarlskog approach for obtaining 
viable relations between the masses of 
the muon and the $s$-quark 
\cite{Chen:2007afa,Chen:2009gf}.
The Cabibbo angle is given by the 
``standard'' expression:
$\theta_c \cong \sqrt{m_d/m_s}$,
$m_d$ and $m_s$ being the masses of the 
$d$- and $s$- quarks. 
As a consequence, in particular, 
of the connection between  $M_{e}$ and  $M_d$, 
the smallest angle in the neutrino 
mixing matrix $\theta_{13}$, 
is related to the Cabibbo angle 
\cite{Chen:2009gf}:
%%%%%%%%%%%%%%%%%%%%%%%%%%%%%%%%%%%%%%%%%%%%%%%%
\be 
\sin\theta_{13} \cong \frac{1}{3\sqrt{2}}\,\sin\theta_{c}\,.
\label{th13}
\ee
%%%%%%%%%%%%%%%%%%%%%%%%%
%
Here we implicitly assumed the ``standard'' 
parametrisation of the PMNS matrix \cite{PDG10}:
 %%%%%%%%%%%%%%%%%%%%%%%%%%%%%%%%%%
\be
\label{eq:Upara}
U = \left( 
\begin{array}{ccc}
c_{12}  \, c_{13} & s_{12}  \, c_{13} & s_{13} e^{-i \delta}  \\[0.2cm] 
-s_{12}  \, c_{23} - c_{12}  \, s_{23}  \, s_{13} e^{i \delta} 
& c_{12}  \, c_{23} -  \, s_{12}  \, s_{23}  \, s_{13} e^{i \delta} 
& s_{23}  \, c_{13}  \\[0.2cm] 
s_{12}  \, s_{23} - c_{12}  \, c_{23}  \, s_{13} e^{i \delta} & 
- c_{12}  \, s_{23} - s_{12}  \, c_{23}  \, s_{13} e^{i \delta} 
& c_{23}  \, c_{13}  
\end{array}   
\right) 
{\rm diag}(1,e^{i \frac{\alpha_{21}}{2}},e^{i \frac{\alpha_{31}}{2}})\,, 
\ee 
%%%%%%%%%%%%%%%%%%%%%%%%%%%%%%%%%
%
where $c_{ij} = \cos\theta_{ij}$ and $s_{ij} = \sin\theta_{ij}$, 
$0\leq \theta_{ij} \leq \pi/2$, 
$\delta$ and  $\alpha_{j1}$, $j=2,3$, are the Dirac and 
the two Majorana CP violation phases \cite{BHP80},
$0\leq \delta \leq 2\pi$ and, in general \cite{EMSPEJP09},
$0 \leq \alpha_{j1}/2 \leq 2\pi$. We will use this 
parametrisation in the discussion which follows.

 The CP violation, 
predicted by the model, can entirely be geometrical 
in origin ~\cite{Chen:2009gf}. This interesting 
aspect of the $SU(5)\times T'$ model we will consider 
is a consequence of one of the 
special properties of the group $T'$, namely, 
that its group theoretical Clebsch-Gordon 
(CG) coefficients are intrinsically complex~\cite{cg}. 
The only dominant source of CP violation 
in the lepton sector of the model 
is the Dirac phase $\delta$.
The two Majorana phases present in the 
PMNS neutrino mixing matrix,  
$\alpha_{21}$ and $\alpha_{31}$,
are predicted to leading order 
to have CP conserving values. 
In the standard parametrisation of
$U_{\rm PMNS}$ 
we have: $\alpha_{21} \cong 0$ and $\alpha_{31} \cong \pi$.
Higher order corrections induce small CP violating 
deviations of the order of few degrees 
from these CP conserving values of the two 
phases \cite{AMSP1109}. 

  The Dirac phase $\delta$ in the 
PMNS matrix is induced effectively by 
the complex CG coefficients of the group $T'$.
As we shall see in Section 2,  $\delta$ can take two 
values in the model considered. 
One was identified in~\cite{Chen:2009gf} and    
is equal approximately to
$\delta \cong 5\pi/4 = 225^{\circ}$,
the precise value being
%%%%%%%%%%%%%%%%%%%%%%%%%%%%%%%%%
\be
\delta \cong 226.9^{\circ}\,.
\label{delta2}
\ee
%%%%%%%%%%%%%%%%%%%%%%%%%%%%%%%
%
 The second possible value of $\delta$ is given
to leading order, as will be discussed in Section 2, by 
%%%%%%%%%%%%%%%%%%%%%%%%%%%%%%%%%
\be
\delta \cong \frac{\pi}{4} = 45^{\circ}\,.
\label{delta3}
\ee
%%%%%%%%%%%%%%%%%%%%%%%%%%%%%%%
%

The tri-bimaximal mixing value of the 
solar neutrino mixing angle $\theta_{12}$, 
which corresponds to $\sin^2\theta_{12} = 1/3$,
is corrected by a quantity which, as it follows 
from the general form of such corrections 
\cite{FPR,HPR07,Marzocca:2011dh}), 
is determined by the angle $\theta_{13}$ 
and the Dirac phase $\delta$:
%%%%%%%%%%%%%%%%%%%%%%%%%
\be 
\sin^2\theta_{12} \cong 
\frac{1}{3} + \frac{2\sqrt{2}}{3}\,\sin\theta_{13}\,\cos\delta 
\label{th120}
\ee
%
%%%%%%%%%%%%%%%%%%%%%%%%
%
In the  $SU(5)\times T'$ model considered, 
$\theta_{13}$ is related to 
the Cabibbo angle, eq. (\ref{th13}).

 The rephasing invariant associated with the 
Dirac phase $\delta$ \cite{CJ85}, $J_{\rm CP}$, 
which determines the magnitude of CP violation 
effects in neutrino oscillations \cite{PKSP3nu88}, 
predicted by the model to leading order 
reads \cite{FPR,HPR07,AMSP1109}: 
%%%%%%%%%%%%%%%%%%%%%%%%%%%%%%%%%%%
\be
J_{CP} \cong \frac{1}{3\sqrt{2}}\sin\theta_{13}\,\sin\delta \cong
\frac{1}{18}\sin\theta_c\,\sin\delta\,. 
% \cong -9.7\times 10^{-3}\,.
\label{JCP}
\ee
%%%%%%%%%%%%%%%%%%%%%%%%%%%%
%

For  $\delta \cong 5\pi/4$, eq. (\ref{delta2}), 
the correction to the TBM value of $\sin^2\theta_{12}$ 
given in eq. (\ref{th120}), is negative and 
$\sin^2\theta_{12} \cong 0.299$, where we have used 
eq. (\ref{th13}) and $\sin\theta_c = 0.22$.
This value lies within the $1\sigma$ allowed range,
found in the global data analysis \cite{Fogli:2011qn}.
We also have, including the higher order corrections 
\cite{Chen:2009gf}: 
$J_{CP} \cong - 9.66\times 10^{-3}$.
If $\delta \cong \frac{\pi}{4}$, eq. (\ref{delta3}), the 
correction to the TBM value of $\sin^2\theta_{12}$ 
is positive and $\sin^2\theta_{12} \cong 0.37$.
According to the analyses performed in \cite{Fogli:2011qn} 
and in \cite{Maltoni1211}, the current neutrino oscillation data   
imply respectively $\sin^2\theta_{12} \ltap 0.36$ 
and $\sin^2\theta_{12} \ltap 0.374$ at 3$\sigma$.
Thus, the case of  $\delta \cong \frac{\pi}{4}$ is 
disfavored by the  data. For the $J_{CP}$ factor 
in this case we get: $J_{CP}  
\cong + 9.95\times 10^{-3}$. 

 Since the neutrino masses, the neutrino 
mixing angle and the CP violating phases 
in the PMNS matrix have essentially fixed
values, the model provides also 
specific predictions \cite{AMSP1109} 
for the sum of the three neutrino masses, 
%%%%%%%%%%%%%%%%%%%%%%%%%%%%%%%%%%%%
\be
m_1 + m_2 + m_3 \cong 5.9\times 10^{-2}~{\rm eV}\,,
\ee
%%%%%%%%%%%%%%%%%%%%%%%%%%%%%%%%%%%
%
as well as for the effective Majorana 
mass in neutrinoless double-beta decay
(see, e.g.,  \cite{BiPet87}):
%%%%%%%%%%%%%%%%%%%%%%%%%%%%%
\be
\meff \cong 3.4\times 10^{-3}~{\rm eV}\,.
\label{meff}
\ee
%%%%%%%%%%%%%%%%%%%%%%%%%%%%%%%%
%

 It should be clear from the preceding 
discussion that the 
$SU(5)\times T^\prime$ model of 
flavour of interest is remarkably 
predictive:
the values of the 
neutrino masses, the type of the neutrino 
mass spectrum, the values of the neutrino 
mixing angles and the CP violating phases in the 
neutrino mixing matrix, as well as 
the effective Majorana mass 
in neutrinoless double beta decay, 
obtained in the model are essentially
free of ambiguities. 
The predictions for $\sin\theta_{13}$, 
$\sin^2\theta_{12}$, $\delta$ and $J_{CP}$ 
can be tested directly in the upcoming 
neutrino oscillation experiments.
The value of $\sin\theta_{13}$ 
one gets in the model,
for instance, is relatively small, $\sin\theta_{13} \cong 0.058$~\footnote{A larger value of $\theta_{13}$ can, in
principle, be obtained along the lines discussed in ref.
\cite{Marzocca:2011dh}.}.
It lies outside the 2$\sigma$, but within the 3$\sigma$, ranges 
of allowed values of  $\sin\theta_{13}$, determined in the global 
analyses of the current neutrino oscillation data 
\cite{Fogli:2011qn,SchwT2V0811}. 
The results of the three reactor 
$\bar{\nu}_e$ experiments on $\theta_{13}$, 
Double Chooz \cite{DCHOOZ}, 
RENO \cite{RENO} and Daya Bay \cite{DayaB}, 
which are currently taking data,  
can provide a critical test of the model.

 In the present article we investigate the prediction of the 
$SU(5)\times T'$ model of flavour proposed in 
\cite{Chen:2007afa,Chen:2009gf} 
for the baryon asymmetry of the Universe. 
The latter is generated in the model via the 
leptogenesis mechanism \cite{FY,kuzmin}.
The dominant source of CP violation in the 
lepton sector and in leptogenesis 
is the Dirac phase $\delta$ 
\footnote{The Casas-Ibarra matrix \cite{Casas:2001sr}, 
which can be an additional source of CP violation 
in leptogenesis, is real in the model 
under discussion \cite{AMSP1109}.}.
Therefore there is a direct connection 
between the baryon asymmetry of 
the Universe and the CP violation 
in neutrino oscillations.

  The generation of the baryon asymmetry in 
the $SU(5)\times T'$ model of interest was 
studied in \cite{MCCKMLG0611}.  
However, the authors of \cite{MCCKMLG0611}
limited the discussion of the 
baryon asymmetry generation to the calculation 
of the CP asymmetries in the additive 
lepton charges, $\epsilon^{\ell}_i$, 
generated in the heavy Majorana neutrino decays, 
$\ell = e,\mu,\tau$, $i=1,2,3$.
They based their conclusions on the results obtained for 
these asymmetries. In the present article 
we perform a complete calculation of the 
baryon asymmetry, i.e., we calculate not only the 
asymmetries  $\epsilon^{\ell}_i$, but also the 
corresponding efficiency factors which 
account for the effects of the CP asymmetry wash-out 
processes, taking place in the Early Universe.
The efficiency factors are computed by solving 
numerically the Boltzmann equations, which describe 
the evolution of the CP violating asymmetries 
in the Early Universe. The results we obtain 
for the lepton asymmetries  $\epsilon^{\ell}_i$
do not agree with those found in \cite{MCCKMLG0611}
and our results for the baryon asymmetry 
contradict the claims made in \cite{MCCKMLG0611}.

%%%%%%%%%%%%%%%%%%%%%%%%%%%%
%
\section{Ingredients}
%
%%%%%%%%%%%%%%%%%%%%%%%%%%
 
 In the $L-R$ convention in which the neutrino 
mass terms are written with the RH 
neutrino fields on the right,
the superpotential of the model leads \cite{Chen:2007afa,Chen:2009gf}
to the following neutrino Dirac mass matrix, 
%%%%%%%%%%%%%%%%%%%%
\begin{equation}
M_{D} = \left( \begin{array}{ccc}
2\xi_{0} + \eta_{0} & -\xi_{0} & -\xi_{0} \\
-\xi_{0} & 2\xi_{0} & -\xi_{0} + \eta_{0} \\
-\xi_{0} & -\xi_{0}+\eta_{0} & 2\xi_{0}
\end{array}\right) \zeta_{0} \zeta^{\prime}_{0} v_{u}
% \equiv h_{D} v_{u} \; ,
\equiv \tilde{Y}_{\nu} v_{u} \;, 
\label{MD}
\end{equation}
%%%%%%%%%%%%%%%%%%%%%%%%%%%%
%
and to the following Majorana mass matrix 
of the RH neutrinos, 
%%%%%%%%%%%%%%%%%%%%%%%%%%%%
\begin{equation}
M_{RR} = \left( \begin{array}{ccc}
1 & 0 & 0 \\
0 & 0 & 1 \\
0 & 1 & 0
\end{array}\right) s_{0} \Lambda \;.
\label{MRR}
\end{equation}
%%%%%%%%%%%%%%%%%%%%%%%%%%%%%%%%
%
In eqs. (\ref{MD}) and (\ref{MRR}),  
$\xi_{0}$, $\eta_{0}$, $\zeta_{0}$, 
$\zeta^{\prime}_{0}$ and $s_{0}$ 
are dimensionless real parameters,
$\Lambda$ is the scale above which 
the $T'$ symmetry is exact,
$\tilde{Y}_{\nu}$ is the matrix of 
the neutrino Yukawa couplings in the 
basis in which the charged lepton and
the RH neutrino mass matrices 
are not diagonal, and $v_{u}$ is the vacuum 
expectation value of the ``up'' Higgs 
doublet field of the SUSY extensions of the 
Standard Model. Thus, the neutrino Dirac mass matrix 
in the model, $M_{D}$, is real and symmetric. 
As can be easily shown, it is 
diagonalised by the tri-bimaximal mixing (TBM) matrix:
%%%%%%%%%%%%%%%%%%%%%%%%%%%%%%%%%%%%%%%%%%
\begin{equation}
U_{TBM} = \left(\begin{array}{ccc}
\sqrt{2/3} & \sqrt{1/3} & 0 \\
-\sqrt{1/6} & \sqrt{1/3} & -\sqrt{1/2} \\
-\sqrt{1/6} & \sqrt{1/3} & \sqrt{1/2}
\end{array}\right) \;.
\label{TBMM}
\end{equation}
%%%%%%%%%%%%%%%%%%%%%%%%%%%%%%%%
%
We have:
%%%%%%%%%%%%%%%%%%%%%%%%%%%%%%%
\begin{equation}
\label{eq:MDdiag}
 U^T_{\mbox{\tiny TBM}}\, M_D\,  U_{\mbox{\tiny TBM}}=
M_{D}^{\mbox{\tiny diag}}\, = diag(3\xi_{0} +
\eta_{0},\eta_{0},3\xi_{0} - \eta_{0})\zeta_0\zeta_0'v_u
\end{equation}
%%%%%%%%%%%%%%%%%%%%%%%%%%%%
%
where all elements in the diagonal matrix $M_{D}^{\mbox{\tiny
diag}}$ are real.

  The RH neutrino Majorana mass matrix 
$M_{RR}$ is diagonalised by the unitary 
matrix $S$:
%%%%%%%%%%%%%%%%%%%%%%%%%%%%%%%%%%%
\begin{equation}
S^T\, M_{RR}\, S = D_N = diag(M_1,M_2,M_3) = s_0\Lambda\,
diag(1,1,1) \,,~~M_j > 0,~~j=1,2,3\,, 
\label{MRRdiag}
\end{equation}
%%%%%%%%%%%%%%%%%%%%%%%%%%%%%%%%%
%
where
%%%%%%%%%%%%%%%%%%%%%%%%%%%%%%%%%%%%%%%%%
\be S= \left(
  \begin{array}{ccc}
    1 & 0 & 0 \\
    0 & 1/\sqrt{2} & -i/\sqrt{2} \\
   0 & 1/\sqrt{2} & i/\sqrt{2} \\
  \end{array}
\right)\,. 
\label{S} 
\ee
%%%%%%%%%%%%%%%%%%%%%%%%%%%%%%%%%%%%%%%%
%
and $M_j$ are the masses of the heavy Majorana neutrinos $N_j$
(possessing definite masses),
%%%%%%%%%%%%%%%%%%%%%%%%%%%%%
\be 
N_j = S^{\dag}_{jl}N_{lR} + S^{T}_{jl}\,C (\bar{N}_{lR})^T =
C(\bar{N}_j)^T\,,~ j=1,2,3\,, 
\label{Nk} 
\ee
%%%%%%%%%%%%%%%%%%%%%%%%%%
%
$C$ being the charge conjugation matrix. 
Thus, to leading order, the
masses of the three heavy Majorana neutrinos $N_j$ coincide, $M_j =
s_0\Lambda \equiv M$, $j=1,2,3$. It follows from eq. (\ref{MRRdiag})
that $S^* S^\dagger$ is a real matrix, so $S^* S^\dagger = SS^T$.

 The effective Majorana mass matrix
of the left-handed (LH) flavour neutrinos, $M_{\nu}$, 
which is generated by the see-saw mechanism,
%%%%%%%%%%%%%%%%%%%%
\begin{equation}
M_{\nu} = -M_{D} M_{RR}^{-1} M_{D}^{T} \; ,
\end{equation}
%%%%%%%%%%%%%%%%%%%%%
%
($\mathcal{L}^{M}_{\nu_L} = 
- \frac{1}{2}\bar{\nu_{L}}\, M_{\nu}\, \nu^c_{R} + h.c.$)
is also diagonalized by the TBM matrix (\ref{TBMM}),
%%%%%%%%%%%%%%%%%%%%
\begin{equation}
% \U_{TBM}^{T} M_{\nu} U_{TBM} =
 U_{TBM}^{T} M_{\nu} U_{TBM} =
\mbox{diag}( (3\xi_{0} + \eta_{0})^{2}, \eta_{0}^{2},
-(-3\xi_{0}+\eta_{0})^{2}) \frac{(\zeta_{0} \zeta_{0}^{\prime}
v_{u})^{2} }{ s_{0}\Lambda} = Q\,diag(m_1,m_2,m_3)\,Q^T\,.
\label{Mnudiag}
\end{equation}
%%%%%%%%%%%%%%%%%
%
Here $Q = i\,diag (1,1,\pm i)$ 
is the matrix which determines, as we shall see,
the leading order 
values of the two Majorana phases in the PMNS matrix,
and $m_k > 0$, $k=1,2,3$, are the masses of 
the three light Majorana neutrinos,
%%%%%%%%%%%%%%%%%%%%%%%%%%%%%%%%%%%%
\be 
m_1 \equiv \frac{(X+3Z)^2}{M}, \quad m_2 \equiv \frac{X^2}{M},
\quad m_3 \equiv \frac{(X-3Z)^2}{M}\,, 
\label{masses} 
\ee
%%%%%%%%%%%%%%%%%%%%%%%%%%%%%%%%%%
%
where $X \equiv \eta_0(\zeta_{0} \zeta_{0}^{\prime} v_{u})$ and $Z
\equiv \xi_0 (\zeta_{0} \zeta_{0}^{\prime} v_{u})$.
In what follows we will ignore the overall unphysical 
factor $i$ in $Q$.
The values of $m_j$ given in eq. (\ref{mj}) correspond to 
\cite{AMSP1109} $X = \pm 1.71 \times 10^{-2} v_u$, and 
$Z = \mp 7.74 \times 10^{-3} v_u $.

 The charged lepton mass matrix $M_e$ is not diagonal; it is 
diagonalised by a bi-unitary transformation: 
$M_e = V_{eR}M_e^{d}U^{\dagger}_{e}$, where
$V_{eR}$ and $U_e$ are unitary matrices and $M_e^{d} =
diag(m_e,m_{\mu},m_{\tau})$, $m_\ell$ being the mass of the charged
lepton $\ell$, $\ell=e,\mu,\tau$. The matrix $U_e$, 
which enters into the expression 
for the PMNS matrix, $U = U^\dagger_eU_{\nu}$, 
diagonalises the matrix $M^{\dagger}_eM_e$.
The charged lepton mass matrix $M_e$  
(with the corresponding mass term 
written in the R-L convention in terms of
the chiral charged lepton fields $l'_{aR}$ and  $l'_{aL}$)
has the following form 
\cite{Chen:2009gf}:
%%%%%%%%%%%%%%%%%%%%%%%%
\begin{eqnarray}
\label{Me}
M_{e} & = & \left( \begin{array}{ccc}
0 & -(1-i) \phi_{0} \psi^{\prime}_{0} & \phi_{0} \psi^{\prime}_{0} \\
(1+i) \phi_{0} \psi^{\prime}_{0} & -3 \psi_{0} \zeta^{\prime}_{0} & \phi_{0} \psi^{\prime}_{0} \\
0 & 0 & \zeta_{0} 
\end{array}\right)\, y_{d}\, v_{d}\, \phi_{0} \;.
\end{eqnarray}
%%%%%%%%%%%%%%%%%%%%%
%
It is related to the 
down-type quark mass matrix $M_d$ via the well-known $SU(5)$ 
relation: $M_e = M^T_d$, with the factor (-3) in $M_e$ 
replaced by 1 in $M_d$. 
The  up-type quark mass matrix in the model has the form 
\cite{Chen:2009gf}:
%%%%%%%%%%%%%%%%%%%%%%%%%%%%
\begin{equation}
M_{u} = \left( \begin{array}{ccc}
i \phi^{\prime 3}_{0}  & (\frac{1-i}{2}) \phi_{0}^{\prime 3} & 0 \\
(\frac{1-i}{2})  \phi_{0}^{\prime 3}  & \phi_{0}^{\prime 3} + 
  (1 - \frac{i}{2}) \phi_{0}^{2} & y^{\prime} \psi_{0} \zeta_{0} \\
0 & y^{\prime} \psi_{0} \zeta_{0} & 1
\end{array} \right)\, y_{t}\,v_{u}, \qquad , 
\label{Mu}
\end{equation}
%%%%%%%%%%%%%%%%%%%%%%%%%%%
%
In eqs. (\ref{Me}) and (\ref{Mu}), 
$\phi_{0}$, $\psi^{\prime}_{0}$,
$\phi^{\prime}_{0}$, $\psi_{0}$, $y_{d}$ and  $y_{t}$ 
are real  dimensionless parameters, 
$y_{d}$ and  $y_{t}$ are Yukawa couplings
and $v_{d}$ is the vacuum expectation value
of the  ``down'' type Higgs doublet
of the SUSY extension of the Standard Model.

 Fitting the quark sector observables and charged 
lepton masses one finds that \cite{Chen:2009gf} 
two of the three angles, present in the ``standard-like'' 
parametrisation of the matrix 
$U_e$, are extremely small, 
%%%%%%%%%%%%%%%%%%%%%%%%%%%%
$ \sin\theta^e_{13} \cong 1.3\times 10^{-5}$,  
$ \sin\theta^e_{23} \cong 1.5\times 10^{-4}$,
%%%%%%%%%%%%%%%%%%%%%%%%%%%%%%
%
while the third satisfies:
%%%%%%%%%%%%%%%%%%%
\be 
\sin\theta^e_{12} = \frac{1}{3}\,\sin\theta_{c}\,.
\label{the12}
\ee
%%%%%%%%%%%%%%%%%%%%
%
Thus, to a very good 
approximation one can set 
$\theta^e_{13}= \theta^e_{23} = 0$,
and in this approximation
$U_e$ takes the form \cite{AMSP1109}: 
$U_e = \Phi\,R_{12}(\theta^e_{12})$, 
where $\Phi = diag(1,e^{i\varphi},1)$ and
%%%%%%%%%%%%%%%%%%%%%%%%%%%%%%%%%%%%%%%%%
\be  
R_{12}(\theta^e_{12}) = 
\left(
  \begin{array}{ccc}
    \cos\theta^e_{12} & \sin\theta^e_{12} & 0 \\
    -\,\sin\theta^e_{12} &  \cos\theta^e_{12}  & 0 \\
   0 & 0 & 1 \\
  \end{array}
\right)\,.
\label{R12}
\ee
%%%%%%%%%%%%%%%%%%%%%%%%%%%%%%%%%%%%%%%%
%

 It follows from the above discussion that 
in the basis in which the charged lepton and the 
RH neutrino mass matrices are diagonal, the 
matrix of neutrino Yukawa couplings $Y_{\nu}$
has the form:
%%%%%%%%%%%%%%%%%%%%%%%%%%
\be 
Y_{\nu} = U^\dagger_e \,\tilde{Y}_{\nu}\, S\,
= \frac{1}{v_u}\,U^\dagger_e \, M_D\, S\,. 
\label{Ynu1}
\ee
%%%%%%%%%%%%%%%%%%%%%%%%%%%%
%
In the same basis, 
the Majorana mass term for the LH flavour
neutrinos, generated by the see-saw mechanism, 
is given by:
%%%%%%%%%%%%%%%%%%%%%%
\be 
M_{\nu} = -\,v^2_u \, Y_{\nu} D_{N}^{-1} Y_{\nu}^{T} =
U\, D_{\nu}\, U^T \;,~ 
\label{MnuYnu} 
\ee
%%%%%%%%%%%%%%%%%%%%%
%
where $D_{N} \equiv diag(M_1,M_2,M_3) = M diag(1,1,1)$,
$D_{\nu} \equiv diag(m_1,m_2,m_3)$  and 
$U$ is the PMNS matrix,
%%%%%%%%%%%%%%%%%%%%%%%%
\be 
U = U^{\dagger}_e U_{TBM} Q\,.
\label{Uprime} 
\ee
%%%%%%%%%%%%%%%%%%%%%%%%%%%%%
%
Using the approximate expression for 
$U_e = \Phi\,R_{12}(\theta^e_{12})$, 
with $\Phi =  diag(1,e^{i\varphi},1)$ and
$R_{12}(\theta^e_{12})$ given by eq. (\ref{R12}),
we get:
%%%%%%%%%%%%%%%%%%%%%%%%%%%%%%%%%%%%%
\begin{eqnarray}
U &\cong& \left(\begin{array}{ccc}
\sqrt{2/3}c^e_{12} + \sqrt{1/6}s^e_{12}e^{-i\varphi} & 
\sqrt{1/3}(c^e_{12} - s^e_{12}e^{-i\varphi}) & 
             \sqrt{1/2}s^e_{12}e^{-i\varphi} \\
  \sqrt{2/3}s^e_{12} - \sqrt{1/6}c^e_{12}e^{-i\varphi} & 
\sqrt{1/3}(s^e_{12} + c^e_{12}e^{-i\varphi}) &
 -\sqrt{1/2} c^e_{12}e^{-i\varphi}\\
-\sqrt{1/6} & \sqrt{1/3} & \sqrt{1/2}
\end{array}\right)\,Q \;,
 \label{PMNSTBM2}
\end{eqnarray}
%%%%%%%%%%%%%%%%%%%%%%%%%%%%%%%%%%%%
%
where $c^e_{12} = \cos\theta^e_{12}$, 
$s^e_{12} = \sin\theta^e_{12}$.

 As was shown in \cite{AMSP1109,Marzocca:2011dh}, 
the phase $\varphi$ in eq. (\ref{PMNSTBM2}) 
and the Dirac phase $\delta$ in eq. (\ref{eq:Upara}) 
are related as follows:
%%%%%%%%%%%%%%%%%%%%%%%%
\be
\delta = \varphi + \pi\,. 
\label{delta}
\ee
%%%%%%%%%%%%%%%%%%%%%
%

  Comparing the expressions in the 
left-hand and right-hand sides of the 
equation $M^{\dagger}_e\,M_e = U_e(M_e^{d})^2U_e^\dagger$
and assuming first following \cite{Chen:2009gf} that 
$\phi_{0} \psi^{\prime}_{0}\psi_{0} \zeta^{\prime}_{0} < 0$ and 
$y^{\prime} \psi_{0} \zeta_{0} < 0$ 
(with $\cos\theta^e_{12}\sin\theta^e_{12} >0$), one finds \cite{AMSP1109}:
%%%%%%%%%%%%%%%%%%%%%%%%
\be
\varphi = \frac{\pi}{4}\,,~~~\delta = \frac{5}{4}\,\pi\,. 
\label{varphi}
\ee
%%%%%%%%%%%%%%%%%%%%%
%
The choice $\phi_{0} \psi^{\prime}_{0}\psi_{0} \zeta^{\prime}_{0} < 0$ 
together with the choice  $y^{\prime} \psi_{0} \zeta_{0} < 0$
(see eq. (\ref{Mu})) allows to get 
the best description of the quark masses and mixing, 
possible in the model considered. 
However, one gets similar
description also in the case of  
$\phi_{0} \psi^{\prime}_{0}\psi_{0} \zeta^{\prime}_{0} > 0$ 
and $y^{\prime} \psi_{0} \zeta_{0} > 0$ 
~\footnote{This observation is based on 
numerical results obtained by M. Spinrath. 
We thank M. Spinrath for communicating to us 
the results of his numerical analysis.}. 
In this latter case we get for $\varphi$ and $\delta$: 
 %%%%%%%%%%%%%%%%%%%%%%%%
\be
\varphi = \frac{\pi}{4}\, \pm \pi\,,~~~\delta = \frac{\pi}{4}\,.
\label{varphi2}
\ee
%%%%%%%%%%%%%%%%%%%%%
%

 Numerically, for $\varphi = \pi/4$ and $s^e_{12} = 0.22/3$ 
(see eq. (\ref{the12})), the PMNS matrix, eq. (\ref{PMNSTBM2}), reads:
%%%%%%%%%%%%%%%%%%%%%%%%%%%%%%%%%%%%
\be
U \simeq\left(
\begin{array}{ccc}
0.836e^{-i1.452^{\circ}} & 0.546e^{i3.139^{\circ}} 
& 0.0518e^{-i45.000^{\circ}} \\
0.367e^{i173.380^{\circ}}  & 0.607e^{i2.829^{\circ}} & -\,0.705 \\
-0.408   &  0.577 & 0.707
\end{array}
\right)\,Q.
\label{PMNSnumeric2}
\ee
%%%%%%%%%%%%%%%%%%%%%%%%%%%%%%%%%%
%
Taking into account the corrections due to
the non-zero values of the angles $\theta^e_{13}$
and $\theta^e_{23}$ in  $U^{\dagger}_e$ on finds 
\cite{Chen:2009gf}: 
%%%%%%%%%%%%%%%%%%%%%%%%%%%%%
%
\be
U \simeq \left(
 \begin{array}{ccc}
  0.838e^{-i1.626^{\circ}}  & 0.543e^{i3.551^{\circ}} &
 0.0582e^{-i45.000^{\circ}} \\
 0.362e^{i172.463^{\circ}}  & 0.610e^{i3.160^{\circ}} &  
 -\,0.705 \\
 -\,0.408  & 0.577 & 0.707
 \end{array}
 \right)\,Q.
 \label{PMNSnumeric1}
 \ee
%%%%%%%%%%%%%%%%%%%%%%%%%%%%%%%%%%%%
%
Obviously, the differences between 
the approximate and the ``exact'' 
matrices (\ref{PMNSnumeric2}) and (\ref{PMNSnumeric1})  
are negligibly small.

  The leading order 
predictions of the 
$SU(5)\times T^\prime$ model for 
$\sin\theta_{13}$,  
$\sin^2\theta_{12}$ and $\sin^2\theta_{23}$
were given in the Introduction 
(see eqs. (\ref{th13}), (\ref{delta2}), 
(\ref{th120}) and the related discussions). 
They can be obtained by comparing 
eqs. (\ref{eq:Upara}) and (\ref{PMNSTBM2})
and using eq. (\ref{the12}).

  Equations (\ref{PMNSnumeric2}) and (\ref{PMNSnumeric1})
allow to determine the values of the Majorana phases 
$\alpha_{21}$ and $\alpha_{31}$. In the parametrisation 
in which the PMNS matrix is written in eqs. (\ref{PMNSTBM2}),
(\ref{PMNSnumeric2}) and (\ref{PMNSnumeric1})
they are fixed by the matrix $Q=diag(1,1,\pm i)$ and read  
$\alpha_{21}/2 = 0$ and $\alpha_{31}/2 = \pi/2$ or $3\pi/2$.
Thus, $\alpha_{21}$ and $\alpha_{31}$ are CP conserving.
Note, however, that the parametrisation 
of the PMNS matrix in eqs.  (\ref{PMNSTBM2}),
(\ref{PMNSnumeric2}) and (\ref{PMNSnumeric1}) 
does not coincide with the standard one. 
Thus, in order to get the values of the 
Dirac and Majorana phases $\delta$ and $\alpha_{21}/2$ 
and $\alpha_{31}/2$ of the standard 
parametrisation of the PMNS matrix, one has to 
bring the expressions (\ref{PMNSnumeric2}) 
or (\ref{PMNSnumeric1}) in a form which 
corresponds to the ``standard'' one 
in eq. (\ref{eq:Upara}). This can be done 
by using the freedom of multiplying the rows of the 
PMNS matrix with arbitrary phases and by shifting 
some of the common phases of the columns to a diagonal 
phase matrix $\tilde{Q}$. The results for the 
``approximate'' and ``exact'' numerical matrices, 
eqs. (\ref{PMNSnumeric2}) and (\ref{PMNSnumeric1}), is:
%%%%%%%%%%%%%%%%%%%%%%%%%%%%%%%%%%%%
\be
U \simeq\left(
\begin{array}{ccc}
0.836 & 0.546 
& 0.0518e^{-i226.69^{\circ}} \\
-\,0.367e^{-i3.48^{\circ}}  & 0.607e^{i1.38^{\circ}} & 0.705 \\
0.408 e^{i3.14^{\circ}}  & -\,0.577e^{-i1.45^{\circ}}  & 0.707
\end{array}
\right)\, \tilde{Q_a}\,Q\,,
\label{PMNSnumeric3}
\ee
%%%%%%%%%%%%%%%%%%%%%%%%%%%%%%%%%%
%
and \cite{AMSP1109}
%%%%%%%%%%%%%%%%%%%%%%%%%%%%%
%
\be
U \simeq \left(
 \begin{array}{ccc}
  0.838  & 0.543 & 0.0582e^{-i226.93^{\circ}} \\
 -\,0.362e^{-i3.99^{\circ}}  & 0.610e^{i1.53^{\circ}} &  0.705 \\
 0.408e^{i3.55^{\circ}}  & -\,0.577e^{-i1.63^{\circ}} & 0.707
 \end{array}
 \right)\,\tilde{Q_e}\, Q\,,
 \label{PMNSnumeric4}
 \ee
%%%%%%%%%%%%%%%%%%%%%%%%%%%%%%%%%%%%
%
where  $\tilde{Q_a} = diag(e^{-i3.14^{\circ}},e^{i1.45^{\circ}},-1)$
and  $\tilde{Q_e} = diag(e^{-i3.55^{\circ}},e^{i1.63^{\circ}},-1)$.
Now comparing eq. (\ref{PMNSnumeric3}) and eq. (\ref{PMNSnumeric4})
with eq. (\ref{eq:Upara}) we can obtain the 
``approximate'' and ``exact'' values of the Dirac and 
the two Majorana phases of the standard parametrisation of the 
PMNS matrix, predicted by the model. 
For the Dirac phase, for instance, 
we find, respectively, $\delta \cong 226.7^{\circ}$
and \cite{Chen:2009gf}  $\delta \cong 226.9^{\circ}$. 
Note that the Majorana phases 
$\alpha_{21}/2$ and $\alpha_{31}/2$
in the standard parametrisation are not CP conserving \cite{AMSP1109}: 
due to the matrix  $\tilde{Q_a}$ (or  $\tilde{Q_e}$) 
they get small CP violating corrections to the CP 
conserving values 0 and $\pi/2$ or $3\pi/2$.
 
 The high precision provided by the
expression (\ref{PMNSTBM2}) 
for the PMNS matrix is more than sufficient 
for the purposes of our investigation and we will use it 
in our further analysis. This allows to get simple analytic 
results for the CP violating asymmetries, relevant 
in leptogenesis, which in turn makes transparent 
and easy to interpret the results we are going to obtain.

 Equation (\ref{MnuYnu}), as is well known, 
allows to express $Y_{\nu}$ in terms of 
$U$, $D_{\nu}$, $D_{N}$ and an 
orthogonal (in general, complex)
matrix \cite{Casas:2001sr} $R$, 
$R^TR=RR^T = {\bf 1}$:
%%%%%%%%%%%%%%%%%%%%%%%%%%%%
\be 
Y_\nu =\frac{1}{v_u}\,
U\,\sqrt{D_{\nu}}\,R\,\sqrt{D_N}\,.
\label{YnuR} 
\ee
%%%%%%%%%%%%%%%%%%%%%%
%
From eqs. (\ref{Ynu1}) -  (\ref{YnuR}) and (\ref{eq:MDdiag}), 
we obtain the following exact 
expression for the matrix $R$:
%%%%%%%%%%%%%%%%%%%%%%%%
\be 
R  =\,(\sqrt{D_{\nu}})^{-1}\, Q^*\, \,M^{diag}_D\, \,U^T_{TBM}\
S\,(\sqrt{D_N})^{-1}\,. 
\label{R1} 
\ee
%%%%%%%%%%%%%%%%%%%%%%%%%%%
%
Using the explicit forms of $Q=diag(1,1,\pm i)$, $M^{diag}_D$,  
$U_{TBM}$, $S$ and $D_N = M\,diag(1,1,1)$ 
we get:
%%%%%%%%%%%%%%%%%%%%%%%%%%%%%%%%%%%%
\be  
R=\left(
\begin{array}{ccc}
 -\sqrt{\frac{2}{3}}  & \frac{ 1}{\sqrt{3}} & 0 \\
 \frac{1}{\sqrt{3}} & \sqrt{\frac{2}{3}}& 0 \\
 0 & 0 & -1
\end{array}
\right)\,.
\label{R2}
\ee 
%%%%%%%%%%%%%%%%%%%%%%%%%%%%%
%
The same expression for the matrix $R$ was 
obtained in \cite{MCCKMLG0611}.
Thus, in the  $SU(5)\times T'$ model 
considered, the R matrix is real, i.e., 
CP conserving \cite{Pascoli:2006ci} (see also 
\cite{EMSTP08}), and symmetric, 
$R^* = R$, $R^T = R$, 
and the elements $R_{k3} = R_{3k} = 0$, 
$k=1,2$.
We note that the signs of
the entries in the 1-2 sector of $R$
depend on the signs of $X$ and $Z$: 
the signs in eq. (\ref{R2}) correspond to 
$X > 0$ and $Z < 0$ 
(see eq. (\ref{masses}) 
and the related comments). 

%%%%%%%%%%%%%%%%%%%%%%
%
\section{Radiatively Induced Leptogenesis}
%
%%%%%%%%%%%%%%%%%%%%%%%%%%%%%%

  As we have seen, the three heavy Majorana neutrinos  
$N_j$ are degenerate in mass at the scale $M_X$ at which the 
Majorana mass matrix of the RH neutrinos is generated.
We will assume that this scale does not exceed the GUT scale, 
$M_{GUT} = 2\times 10^{16}$ GeV: $M_X \leq M_{GUT}$. 
Actually, in the SUSY $SU(5)\times T^{\prime}$ model considered,
we have $M_X = M_{GUT}$. 
Given the fact that the 
$R$ matrix is real and CP conserving, 
the baryon asymmetry can only be generated in 
the regime of  flavoured leptogenesis 
\cite{Barbieri99,davidsonetal}. 
The regimes of 2-flavour 
and 3-flavour leptogenesis are realised, in general, 
for values of the masses $M_j \cong M$, $j=1,2,3$, 
of the heavy Majorana neutrinos satisfying 
\cite{Pascoli:2006ci} 
$M\ltap T < (1 + \tan^2\beta)\times 10^{12}$ GeV and 
$M\ltap T < (1 + \tan^2\beta)\times 10^{9}$ GeV, respectively, 
where $T$ is the temperature of the Early Universe 
and $\tan\beta = v_u/v_d$ is the ratio of the vacuum expectation 
values of the two Higgs doublet fields, 
present in the SUSY theories, 
$v \equiv \sqrt{v^2_u + v^2_d} = 174$ GeV.
If the heavy Majorana neutrinos would be degenerate 
in mass at the scale (temperatures) at which the 
flavoured leptogenesis can take place, as is well known,
no net baryon asymmetry would be generated.
However, if leptogenesis takes place 
at a scale $M_{FLG} < (\ll) M_X$, 
higher order corrections accounted for by the 
renormalisation group (RG) equations describing 
the change of the masses $M_j$ 
with the change of the energy scale from $M_{X}$
to $M_{\rm FLG} \ltap  (1 + \tan^2\beta)\times 10^{12}$ GeV, 
lift the degeneracy of $N_j$ \cite{Pokorski,Casas:1999ac,GF03,KT04}, 
generating relatively small splittings between $M_1$, $M_2$  
and $M_3$: $\Delta M_{ij}(M_{\rm FLG}) \equiv 
M_i(M_{\rm FLG}) - M_j(M_{\rm FLG}) \neq 0$, $i\neq j=1,2,3$. 
Since the mass splittings 
$|\Delta M_{ij}(M_{\rm FLG})|$ thus 
generated are exceedingly small, we expect the 
baryon asymmetry to be generated in the regime 
of resonant flavoured leptogenesis
\cite{Pilaf97,Pascoli:2006ci}.

  In the case of resonant flavoured leptogenesis, 
the CP violating asymmetry 
in the lepton charge $L_l$, $l=e,\mu,\tau$, 
generated in the out of equilibrium decays of the heavy Majorana
neutrino $N_j$ taking place at the scale 
$M_{FLG}$, is given by \cite{Pascoli:2006ci}: 
%%%%%%%%%%%%%%%%%%%%%%%%%%%%%%%%%%
\be 
\epsilon^{\ell}_i \equiv\frac{\Gamma(N_i\rightarrow \ell^-\,H^+) 
+ \Gamma(N_i\rightarrow \nu_{\ell}\,H^0)
- \Gamma(N_i \rightarrow \ell^+H^-)  
- \Gamma(N_i\rightarrow \bar{\nu}_{\ell}\,\bar{H}^0)}
{\Gamma(N_i\rightarrow \ell^-\,H^+) + \Gamma(N_i\rightarrow \nu_{\ell}\,H^0)+
\Gamma(N_i \rightarrow \ell^+H^-) + 
\Gamma(N_i\rightarrow \bar{\nu}_{\ell}\,\bar{H}^0)} = 
-\, \frac{1}{8\pi}\sum_{j\neq i} S_{ij}\mathcal{I}^{\ell}_{ij}\,.
\label{elj1}
\ee
%%%%%%%%%%%%%%%%%%%%%%%%%%%%%%%
%
Here
%%%%%%%%%%%%%%%%%%%%%%%%%%%%%%%%
\begin{gather} 
S_{ij}= \frac{M_i M_j \Delta M^2_{ji}}{(\Delta M^2_{ji})^2+M_i^2
\Gamma_j^2}\,,
\qquad \mathcal{I}^{\ell}_{ij}=
\frac{Im[(Y_{\nu}^\dagger Y_{\nu})_{ij}
(Y_{\nu})^\ast_{\ell, i}(Y_{\nu})_{\ell, j}]}
{(Y_{\nu}^\dagger Y_{\nu})_{ii}}\,.
\label{lepto}
\end{gather}
%%%%%%%%%%%%%%%%%%%%%%%%%%%%%%%%
%
where $Y_{\nu}$ is defined in  eqs. (\ref{YnuR}),
%%%%%%%%%%%%%%%%%%%%%%%%%%%%%%
\be
\Gamma_j= \frac{1}{8
\pi}(Y_{\nu}^\dagger Y_{\nu})_{jj}M_j\,,
\label{gamma}
\ee
%%%%%%%%%%%%%%%%%%%%%%%%%%%%%%
% 
and
%%%%%%%%%%%%%%%%%%%%%%%%%%%%%%%%
\be 
\Delta M^2_{ji}\equiv M_j^2 -M_i^2\cong 2M_i^2 \delta^{N}_{ji},
\qquad  \delta^{N}_{ji}= \frac{M_j}{M_i}-1\,,~~j\neq i\,.
\label{DM2dM}
\ee
%%%%%%%%%%%%%%%%%%%%%%%%%%%%%%%%
%
The parameter $\delta^{N}_{ji}$ describes the 
deviation from complete degeneracy of the masses of the 
the heavy Majorana neutrinos $N_j$ and $N_i$.
All quantities which appear in eqs. (\ref{elj1}) - (\ref{DM2dM})
should be evaluated at the leptogenesis scale $M_{FLG}$.
The baryon asymmetry is generated in the regime of 
resonant leptogenesis if at $M_{FLG}$ the
following condition is fulfilled:
%%%%%%%%%%%%%%%%%%%%%%%%%%%%%%%%
\be 
M_i\Gamma_j \cong \Delta M^2_{ji}\,,~~~i\neq j\,.
\label{ResLGCon}
\ee
%%%%%%%%%%%%%%%%%%%%%%%%%%%%%%%%
%

 We have discussed above the asymmetry generated 
in the decays of the heavy Majorana neutrinos $N_i$ 
into the Higgs and lepton doublets.
A  lepton flavour asymmetry $\epsilon^{\tilde{\ell}}_i$ is also 
generated from the out-of-equilibrium decays of $N_i$
in the Higgsino and slepton doublets $\tilde{\ell}$. Similarly,  
the sneutrinos $\tilde{N}_i$ generate  CP asymmetries 
$\epsilon^{\ell}_{\tilde{i}}$ and
$\epsilon^{\tilde{\ell}}_{\tilde{i}}$ with, respectively, 
$\ell$ and $\tilde{\ell}$ in the final state.
As can be shown, one has neglecting soft SUSY breaking terms: 
$\epsilon^{\ell}_i=\epsilon^{\tilde{\ell}}_i=
\epsilon^{\ell}_{\tilde{i}} = \epsilon^{\tilde{\ell}}_{\tilde{i}}$.

  It follows from eq. (\ref{lepto}) that the 
necessary conditions for a successful resonant flavoured
leptogenesis include:
i) the presence of  CP violating phases in the matrix 
of neutrino Yukawa couplings $Y_{\nu}$;
ii) non-vanishing off-diagonal elements of the matrix  
$Y_{\nu}^\dagger Y_{\nu}$:
$(Y_{\nu}^\dagger Y_{\nu})_{ij} \neq 0$ for $i\neq j$;
iii) non-degeneracy of the heavy Majorana neutrino 
masses $M_i$: $\delta^{N}_{ji} \neq 0$,  $i\neq j$.
The first requirement is fulfilled by the 
presence of the CP violating phases in 
the neutrino mixing matrix $U$. 
The second and third general requirements are
satisfied, as we are going to discuss next, 
owing to the RG corrections 
in the quantities $M_i$ and $Y_{\nu}$, 
which have to be included when the latter 
are evaluated at the leptogenesis scale $M_{FLG}$. 

 The RG running of the heavy Majorana 
neutrino masses $M_i$ depends on the quantity 
$Y_{\nu}^\dagger Y_{\nu}$ \cite{Casas:1999ac}.
It proves convenient to work at the scale 
$M_{X}$ in a basis of the heavy 
Majorana neutrino fields in which 
the matrix $Y_{\nu}^\dagger Y_{\nu}$ is diagonal.
This can be achieved by  
performing an orthogonal transformation of 
$N_j$. The latter can be done without affecting the 
heavy Majorana neutrino mass term 
since at the scale of interest the 
heavy Majorana neutrinos $N_j$ 
are degenerate in mass. The change of basis, 
$N_j =  O^T_{jk} N'_k$, where $O$ is an 
orthogonal matrix, implies the following change of 
the matrix of neutrino Yukawa couplings: 
$Y^\prime_\nu =  Y_\nu O$.
Using eq. (\ref{YnuR}) and the facts that 
$D_N = M\,diag(1,1,1)$ and the matrix $R$ is 
real and orthogonal, there always exists 
an orthogonal matrix $O$ such that 
$RO$, and correspondingly 
 $Y_{\nu}^{\prime\dagger} Y^\prime_{\nu}$, are 
diagonal matrices. 
Taking into account the explicit 
form of the matrix $R$ in the model 
considered, eq. (\ref{R2}), 
in what follows we will use 
%%%%%%%%%%%%%%%%%%%%%%%%%%%%%%%%%%%%%
\be
 O\equiv \left(
\begin{array}{ccc}
 \cos\theta  & \sin\theta  & 0 \\
 -\sin\theta & \cos\theta& 0 \\
 0 & 0 & 1
\end{array}
\right)=\left(
\begin{array}{ccc}
 \sqrt{\frac{2}{3}}  & \frac{ 1}{\sqrt{3}} & 0 \\
 -\frac{1}{\sqrt{3}} & \sqrt{\frac{2}{3}}& 0 \\
 0 & 0 & 1
\end{array}
\right). 
\label{YnuR1}
\ee
%%%%%%%%%%%%%%%%%%%%%%%%%%%%%%%%%
%
It is easy to verify that $RO = diag(-1,1,-1)\equiv I$,
$I_{jj} = \eta_j$, with $\eta_2 = -\,\eta_{1,3} = 1$. 
The matrix of neutrino Yukawa couplings $Y^\prime_{\nu}$
is given by:
%%%%%%%%%%%%%%%%%%%%%%%%%%%%%%%%%%%
\ba 
Y^\prime_\nu\equiv Y_\nu O =\frac{\sqrt{M}}{v_u}U\,\sqrt{D_\nu} I\,,
\label{Yprime}
\ea
%%%%%%%%%%%%%%%%%%%%%%%%%%%%%%%%%%%%
%
In the new basis we have 
$(Y_{\nu}^{\prime\dagger} Y'_{\nu})_{ij}=0$ for $i\neq j$, 
and
%%%%%%%%%%%%%%%%%%%%%%%%%%%
\be 
(Y_{\nu}^{\prime\dagger} Y'_{\nu})_{ii} = \,\frac{M_i}{v_u^2}\,m_i,
\qquad \Gamma'_i= \frac{M_i^2}{8 \pi v_u^2}\,m_i\,,~~~~~M_i \cong M\,,
\label{Y'Y'}
\ee 
%%%%%%%%%%%%%%%%%%%%%%%%%%
%
where $\Gamma^\prime_i$ is the $N'_i$ total decay width.

  The expression for the CP violating asymmetry
$\epsilon^{\ell}_i$ in the new basis in which 
$Y_{\nu}^{\prime\dagger} Y'_{\nu}$ is diagonal at $M_X$
can be obtained from eqs. (\ref{elj1}) - (\ref{gamma}) 
by replacing  $Y_{\nu}$ and  $\Gamma_i$
with  $Y'_{\nu}$ and  $\Gamma^\prime_i$, respectively.
Note, however, that in the new basis we have 
$\mathcal{I}^{\ell}_{ij}=0$. Thus, the CP 
violating asymmetries $\epsilon^{\ell}_i$ will be 
zero unless non-diagonal elements 
of $Y_{\nu}^{\prime\dagger} Y'_{\nu}$ 
are radiatively generated at 
the leptogenesis scale $M_{FLG} < (\ll) M_X$.

 As will be shown later, in the model considered 
a non-zero baryon asymmetry can be produced only in 
the regime of 3-flavoured leptogenesis, i.e. 
for $M < (1 + \tan^2\beta)\times 10^{9}~{\rm GeV} 
\ltap 4.9\times 10^{12}$ GeV, where we have used 
the constraint $\tan \beta \ltap 70$  
(see
\footnote{In the calculation of the baryon asymmetry 
we will values of  $\tan \beta \sim 10$, which are 
much smaller than the quoted maximal value.}, 
e.g., \cite{10041993}).
Taking into account that $m_i \ltap 5\times 10^{-2}$ eV 
and $v = 174$ GeV, we get 
$|(Y_{\nu}^{\prime\dagger} Y'_{\nu})_{ii}| \ltap 8\times 10^{-3} \ll 1$.

  In the new basis, the running of 
the heavy Majorana neutrino 
masses $M_i$ is governed by the following equation 
\cite{Casas:1999ac}:
%%%%%%%%%%%%%%%%%%%%%%%%%%%%%%%%
\begin{gather} 
\frac{dM_i}{dt}=
 4\, (Y_{\nu}^{\prime\dagger} Y_{\nu}^\prime)_{ii}\, M_i\,,
\quad t\equiv \frac{1}{16 \pi^2}\ln \frac{\mu}{M_{X}} \,,
\label{RGM1}
\end{gather}
%%%%%%%%%%%%%%%%%%%%%%%%%%%%%%%%
%
where the initial conditions are at 
the scale $\mu = M_{X}$ at which
$M_i = M$, $i=1,2,3$, and the masses 
$M_i$ are evaluated at the scale 
$\mu = M_{FLG} < (\ll) M_{X}$. 
The latter coincides, 
up to negligibly small corrections,
with $M$: $M_{FLG} \cong M$.
The running of the masses $M_i$ 
from  $M_{X}$ to $M_{FLG} \cong M$
induces the splitting 
between the masses of the heavy
Majorana neutrinos, necessary for a 
potentially successful leptogenesis.
The solutions of the equations (\ref{RGM1}) 
\cite{GF03,KT04} lead 
for $|(Y_{\nu}^{\prime\dagger} Y_{\nu}^\prime)_{ii}| \ll 1$
to the following expression for the 
mass splitting parameter $\delta^{N}_{ji}$:
%%%%%%%%%%%%%%%%%%%%%%%%%%%%%%%%
\be 
\delta^{N}_{ji} \cong - \, 4 [(Y_{\nu}^{\prime\dagger}
Y'_{\nu})_{jj}-(Y_{\nu}^{\prime\dagger} Y'_{\nu})_{ii}]\tilde{t} 
\cong -\, 4\, \frac{M}{v_u^2}(m_j - m_i)\,\tilde{t}\,,~j\neq i\,, \qquad
\tilde{t}=\frac{1}{16 \pi^2}\ln\left(\frac{M_{X}}{M}\right)\,.
\label{deltaNji01}
\ee
%%%%%%%%%%%%%%%%%%%%%%%%%%%%%%%%
% 
For $M_X/M = 2\times 10^6$,  $2\times 10^5$ and $2\times 10^4$,
we get $\tilde{t}=0.092$, $0.077$, $0.063$. 
The corresponding values of $\delta^{N}_{ji}$ 
are given in Table \ref{tab:delta}.
%%%%%%%%%%%%%%%%%%%%%%%%%%%%%%%%%%%%%%%%%%%%%%%%
\begin{table}[h!] \centering \vspace{15pt}
\begin{tabular}{|l|c|c|c|}
\hline
& $M_X/M =2\times 10^{6}$ & $M_X/M =2\times 10^{5}$ & $M_X/M=2\times 10^{4}$ \\
\hline\hline
$\delta^{N}_{21}$ & -9.28 $\times10^{-7}$ & -7.81 $\times10^{-6}$ & -6.33$\times 10^{-5}$\\
\hline
$\delta^{N}_{31}$ & -5.80$\times 10^{-6}$ & -4.88 $\times10^{-5}$ & -3.96 $\times 10^{-4}$ \\
\hline
$\delta^{N}_{32}$ & -4.87$\times10^{-6}$& -4.10 $\times10^{-5}$& -3.33 $\times 10^{-4}$\\\hline
\end{tabular}
\caption{ Values of the heavy Majorana mass splitting 
parameter $\delta^{N}_{ij}$.
\label{tab:delta}
}
\end{table}
%%%%%%%%%%%%%%%%%%%%%%%%%%%%%%%%%%%%%%%%
%

  The elements of the matrix of neutrino Yukawa couplings 
$Y'_{\nu}$ also evolve with the scale $\mu$ when the latter
diminishes from $M_{X}$ to $M_{FLG} \cong M$. 
This change is governed by the RG equations for 
$(Y'_{\nu})_{\ell i}$, whose general form was given 
in \cite{Pokorski,Casas:1999ac,GF03,KT04}. 
In the case considered by us we have at $M_{X}$:
$(Y_{\nu}^{\prime\dagger} Y'_{\nu})_{ij}=0$, $i\neq j$, and 
\footnote{For a matrix of neutrino Yukawa couplings $Y_{\nu}$ such that
${\rm Re}[(Y_{\nu}^{\dagger} Y_{\nu})_{ij}] \neq 0$, $i\neq j$, 
the RG equations for $Y_{\nu}$
have a singularity in the case of 
degenerate in mass heavy Majorana neutrinos
\cite{Pokorski,GF03,KT04}.
As a consequence, the quantity 
$(Y_{\nu}^{\dagger} Y_{\nu})_{ij}$, $i\neq j$, 
that enters into the expression for 
the CP violating asymmetry $\epsilon^{\ell}_i$,
does not vary continuously with the scale 
when the latter changes from $M_X$ to
$M_{FLG}$. This fact was not taken into account 
in the calculation of the asymmetries 
$\epsilon^{\ell}_i$ performed in \cite{MCCKMLG0611}.
Since in the basis in which we work we have 
$(Y_{\nu}^{\prime\dagger} Y'_{\nu})_{ij}=0$, $i\neq j$, 
at $M_{X}$ at which $M_j = M$, $j=1,2,3$, 
the indicated problem does not appear 
when we consider the RG evolution of 
$(Y'_{\nu})_{\ell i}$ and of 
$(Y_{\nu}^{\prime\dagger} Y'_{\nu})_{ij}$.
}
$|(Y_{\nu}^{\prime\dagger} Y'_{\nu})_{ii}|\ltap 8\times 10^{-3}$.
In this case the RG equations 
for $(Y'_{\nu})_{\ell i}$ \cite{KT04} simplify 
considerably and read:
%%%%%%%%%%%%%%%%%%%%%%%%%%%%%%%%%%%%%
\be
\frac{d (Y^\prime_\nu)_{\ell i}}{dt} \cong
\left[ 3\sum_{q=u,c,t} y^2_q  -\,\frac{3}{5}\,g_1^2 - 3\,g_2^2
+ y^2_{\ell} \right]\,(Y^\prime_\nu)_{\ell i}\,,
~~\ell=e,\mu,\tau\,,~~i=1,2,3\,,
\label{RGY1} 
\ee
%%%%%%%%%%%%%%%%%%%%%%%%%%%%%%%%
%
where 
$y_{q}$, $q=u,c,t$, and  $y_{\ell}$, $\ell=e,\mu,\tau$, 
are the charge 2/3 quark and charged 
lepton Yukawa couplings, $g_{\,1,2}$ are the $U(1)_Y$ 
and $SU(2)$ gauge couplings of the 
Standard Model and we have neglected terms 
$\propto Y_{\nu}^{\prime\dagger} Y'_{\nu}$.
The quantities which appear in the 
square brackets in the r.h.s. of 
eq. (\ref{RGY1}) evolve with the scale 
$\mu$ as it decreases from $M_X$, but the effects 
of their evolution are subdominant for the problem 
under study and we will neglect them.
Thus, we will use their values at the scale $M_X$, 
 which will be assumed to be close, or equal, 
to $M_{GUT}$.

 We are interested in the quantities
$(Y_{\nu}^{\prime\dagger} Y'_{\nu})_{ij}$, $i\neq j$, 
at the scale $M_{FLG} \cong M$, 
which enter into the expression for the 
CP violating asymmetry $\epsilon^{\ell}_i$. 
Since these quantities are zero at $M_X$, they 
can get non-zero values at $M_{FLG}$ 
due only to the term involving the charged lepton 
Yukawa coupling $y^2_{\ell}$ in the RG equation 
(\ref{RGY1}) \cite{GF03,KT04}. The solutions of the RG 
equations (\ref{RGY1}) 
in the leading logarithmic approximation
lead to the following result:
%%%%%%%%%%%%%%%%%%%%%%%%%%%%%%%%%
\be 
(Y_{\nu}^{\prime\dagger} Y^\prime_{\nu})_{ij}
\cong 
-\, 2\,y_\tau^2\,
(Y_{\nu}^{\prime\ast})_{\tau i}(Y^\prime_{\nu})_{\tau j}\, 
\widetilde{t}\,,~~~~~i\neq j\,. 
\label{YYatM1}
\ee 
%%%%%%%%%%%%%%%%%%%%%%%%%%%%%%%%%
%
where $y_\tau= (m_\tau/v_d)\cong (m_\tau/v)\sqrt{1 + \tan^2\beta}$
is the $\tau$ Yukawa coupling, $m_\tau$ being the $\tau$ mass,  
and $v = \sqrt{v^2_u + v^2_d} = 174$ GeV.  
Neglecting relatively small effects, the 
quantities in the r.h.s. of eq. (\ref{YYatM1})
can be taken at the scale $M_X$.
Note that even though  at $M_{X}$
the off-diagonal elements of 
$Y_{\nu}^{\prime\dagger} Y'_{\nu}$
are zero, they have non-zero values at the 
leptogenesis scale $M_{FLG}$ due to the radiative
corrections. 

 Using the result obtained for 
$(Y_{\nu}^{\prime\dagger} Y^\prime_{\nu})_{ij}$, 
eq. (\ref{YYatM1}), and eqs. (\ref{elj1}),  (\ref{Yprime}) 
and (\ref{Y'Y'}), we get 
for the CP violating asymmetry:
%%%%%%%%%%%%%%%%%%%%%%%%%%
\ba 
\epsilon^{\ell}_i 
=\, +\, \frac{1}{8\pi}\,y_\tau^2\,\widetilde{t}\, 
\sum_{i\neq j}\,\frac{\delta^N_{ji}}
{\left[ (\delta^N_{ji})^2 + 
\left( \dfrac{M_j m_j}{16 \pi v_u^2}\right)^2\right]}\, 
\frac{M_j m_j}{v_u^2}\,
{\rm Im}\, \left[ U^{\ast}_{\tau i}\,U_{\tau j}\,
U^{\ast}_{\ell i}\,U_{\ell j} \right]\,.
\label{leptof}
\ea
%%%%%%%%%%%%%%%%%%%%%%%
%

 It follows from the expression (\ref{leptof}) 
for $\epsilon^{\ell}_i$
we have derived that $\epsilon^{e}_i + \epsilon^{\mu}_i + 
\epsilon^{\tau}_i =0$, $i=1,2,3$. This result is a 
consequence of the fact that 
the $R$ matrix in the model considered is 
CP conserving (see, e.g., \cite{Pascoli:2006ci}). One can 
easily convince oneself using the explicit expression 
for the PMNS matrix (\ref{PMNSTBM2})
that we also have: $\epsilon^{\tau}_i =0$, $i=1,2,3$.
The same conclusion is reached also if one uses the 
PMNS matrix in which the higher order corrections 
have been included~\footnote{It is claimed in \cite{MCCKMLG0611} that 
$\epsilon^{\tau}_i \neq 0$, which does not correspond to 
the result $\epsilon^{\tau}_i = 0$ we obtain. 
The latter is not difficult to verify.}, 
eq. (\ref{PMNSnumeric1}) or (\ref{PMNSnumeric4}).
Thus, in the SUSY $SU(5)\times T^{\prime}$
model of interest the baryon asymmetry 
can be generated only in the regime 
of 3-flavoured leptogenesis 
\cite{davidsonetal}.

  The requirement that the baryon asymmetry is generated 
in the 3-flavoured thermal leptogenesis regime combined with
the upper limit on $\tan\beta$ implies: 
$M \ltap 4.9\times 10^{12}$ GeV. As is not difficult to show, 
we have for $M \ltap 10^{13}$ GeV:
%%%%%%%%%%%%%%%%%%%%%
\be 
(\delta^N_{ji})^2  \gg 
\left (\frac{1}{16 \pi}\frac{M_j m_j}{ v_u^2}\right)^2\,. 
\ee
%%%%%%%%%%%%%%%%%%%%%%%%%%%%%
%
For  $M = 10^{13}$ GeV, $(\delta^N_{ji})^2$ is bigger by a factor 
of 10 than the term in the right-hand side of the above inequality.
Neglecting the correction due to the latter, we get a rather 
simple expression for the asymmetry $\epsilon^{\ell}_i$:
%%%%%%%%%%%%%%%%%%%%%
\ba
\label{epsf1}
 \epsilon^{\ell}_i  & \cong &- \,\frac{ y_\tau^2}{32\pi}\sum_{i\neq
j} \frac{m_j}{m_j-m_i} {\rm Im} \left[ U^{\ast}_{\tau i}\,U_{\tau j}\,
U^{\ast}_{\ell i}\, U_{\ell j}\right]\,,
\ea
%%%%%%%%%%%%%%%%%%%%%%%%
%
where we have used eqs. 
(\ref{deltaNji01}) and (\ref{leptof}).

  Expression (\ref{epsf1}) 
for  $\epsilon^{\ell}_i$ does not depend explicitly 
on the masses of the heavy Majorana neutrinos and 
on the RG factor $\widetilde{t}$.
Thus, the CP-asymmetries  $\epsilon^{\ell}_i$ are entirely 
determined by the $\tau$ Yukawa coupling and the 
low-energy neutrino mixing parameters, 
i.e., the neutrino masses, 
the neutrino mixing angles and CP violating phases in the 
neutrino mixing matrix. They depend weakly on scales $M_X$ and $M$, 
e.g., via the running of the $\tau$ Yukawa coupling. 
The asymmetries  $\epsilon^{\ell}_i$ depend quadratically on the $\tau$ 
Yukawa coupling and thus on $\tan^2\beta$. 
This dependence is crucial for having a viable thermal 
leptogenesis in the $SU(5)\times T'$ model of flavour 
under consideration.

From eq. (\ref{epsf1}), using eqs. (\ref{PMNSTBM2}) and (\ref{JCP}), 
we obtain: 
%%%%%%%%%%%%%%%%%%%%%
\ba
\label{eif1}
 \epsilon^{e}_i  & \cong &- \,\frac{ y_\tau^2}{32\pi}\, J_{CP}\, 
\sum_{j\neq i} 
\frac{m_j}{m_j-m_i}\,\rho_{ji}\,,~~\epsilon^{\mu}_i = -\epsilon^{e}_i\,,
~i=1,2,3\,,   
\ea
%%%%%%%%%%%%%%%%%%%%%%%%
%
where $\rho_{ji} = - \rho_{ij}$, $i\neq j$, 
and  $\rho_{21} = \rho_{31} =  \rho_{23} = +1$.
Thus, the CP violating asymmetries $\epsilon^{e,\mu}_i$, $i=1,2,3$, 
are all proportional to the $J_{CP}$ factor, which 
determines the magnitude of CP violation effects in the 
flavour neutrino oscillations. 

   The final lepton number asymmetry, which is partially 
converted into a non-zero baryon number asymmetry 
by the fast sphaleron interactions in the thermal bath 
in the Early Universe, receives a contribution from 
the out-of-equilibrium 
decays of the three heavy Majorana neutrinos (sneutrinos)
 $N^{\prime}_i$ ($\tilde{N^\prime}_i$), which are quasi-degenerate in mass. 
The amount of matter-antimatter asymmetry
predicted by the model is computed numerically 
by solving the corresponding system of Boltzmann equations.
We report below  the relevant set of Boltzmann equations in supersymmetric leptogenesis \cite{Antusch:2006cw,Fong:2010qh}
for the lepton flavour (lepton charge) asymmetries 
$\hat{Y}_{\Delta_{\ell}}\equiv Y_{\Delta_\ell}+ Y_{\Delta_{\tilde{\ell}}}$, 
with $\Delta_{\ell(\tilde{\ell})}\equiv B/3-L_{\ell(\tilde{\ell})}$:
~\footnote{ As was pointed out earlier, the CP asymmetries 
$\epsilon_i^\tau$ ($i=1,2,3$) are equal to zero in the model 
we are discussing. Nonetheless, a source term for $\Delta_{\tau(\tilde{\tau})}$
is provided by  non-zero $\hat{Y}_{\Delta_{e,\mu}}$, as is explicit from the
flavoured Boltzmann equation (\ref{YDeltaell}).
} 
%%%%%%%%%%%%%%%%%%%%%%%%%%%%%%%%
\begin{eqnarray}
	\frac{d Y_{N^\prime_i}}{dz} & = & 
- \frac{z}{s H(M_{FLG})}\,\,2\,\left(\gamma_{D}^i\,+ 
\,\gamma_{S,\,\Delta L=1}^i  \right)\,
\left(\frac{Y_{N^\prime_i}}{Y^{\rm eq}_{N^\prime_i}}\,-1\right)\,,
\label{YNNBE}\\
	\frac{d Y_{\widetilde{N^\prime}_i}}{dz} & = & 
- \frac{z}{s H(M_{FLG})}\,2\,\left(\gamma_{D}^{\tilde{i}}\,+ 
\,\gamma_{S,\,\Delta L=1}^{\tilde{i}}  \right)
\,\left(\frac{Y_{\widetilde{N^\prime}_i}}{Y^{\rm eq}_{\widetilde{N^\prime}_i}}\,-1\right)\,,\\
\frac{d\, \hat{Y}_{\Delta_\ell}}{dz} & = & - \frac{z}{s H(M_{FLG})}\,\sum\limits_{i=1}^3\,\left[\, \left(\epsilon_{i}^\ell+\epsilon_{i}^{\tilde{\ell}}\right)\left(\gamma_{D}^i+\gamma_{S,\,\Delta L=1}^i  \right)
\left(\frac{Y_{N^\prime_i}}{Y^{\rm eq}_{N^\prime_i}}-1\right)+ 
\left(\epsilon_{\tilde{i}}^\ell+\epsilon_{\tilde{i}}^{\tilde{\ell}}\right)\left(\gamma_{D}^{\tilde{i}}+\gamma_{S,\,\Delta L=1}^{\tilde{i}}  \right)
	\left(\frac{Y_{\widetilde{N^\prime}_i}}
{Y^{\rm eq}_{\widetilde{N^\prime}_i}}-1\right)\, \right.\nonumber\\
&& \left.-\left(\frac{\gamma_D^{i,\ell}\,+\,\gamma_D^{i,\tilde{\ell}}}{2}
	\,+\,\gamma_{W,\,\Delta L=1}^{i,\ell}\,+\,\gamma_{W,\,\Delta L=1}^{i,\tilde{\ell}}\,+\,\frac{\gamma_D^{\tilde{i},\ell}\,+\,\gamma_D^{\tilde{i},\tilde{\ell}}}{2}
	\,+\,\gamma_{W,\,\Delta L=1}^{\tilde{i},\ell}\,+\,\gamma_{W,\,\Delta L=1}^{\tilde{i},\tilde{\ell}}   \right)
	\,\frac{\sum_{\ell^\prime}A_{\ell\ell^\prime}\,\hat{Y}_{\Delta_{\ell^\prime}}}{\hat{Y}_\ell^{\rm eq}}\,\right]\,. \label{YDeltaell}
\end{eqnarray}
%%%%%%%%%%%%%%%%%%%%%%%%%%%%%%%
%
Here  $Y_{N^\prime_i}$ ($Y^{\rm eq}_{N^\prime_i}$) is the 
$N^\prime_i$ ($N^\prime_i$-equilibrium) 
% number density, 
abundance, $z\equiv M_{FLG}/T$, $T$ being the temperature of the thermal
bath, 
$s$ is the entropy density and $H(T)$ is the expansion rate of the Universe. 
The quantity $\gamma_{D}^{i}$  ($i=1,2,3$) is the thermally averaged 
total decay rate of the Majorana neutrino $N^\prime_{i}$ 
into the SM lepton and Higgs
doublets. Similarly, $\gamma^{i}_{S,\,\Delta L=1}$ is 
the corresponding $\Delta L=1$ thermal scattering rate of $N^\prime_{i}$ 
with SM leptons, quarks and gauge bosons. The flavour 
dependent washout processes involving $N^\prime_{i}$ inverse decays and 
the relative  $\Delta L=1$ scatterings are denoted as  
$\gamma_{D}^{i,\ell(\tilde{\ell})}$  and  
$\gamma^{i,\ell(\tilde{\ell})}_{W,\,\Delta L=1}$, 
respectively. Finally, the matrix elements of $A$ in 
supersymmetric type I see-saw scenarios are \cite{Fong:2010qh}: 
$A_{\alpha\beta}=16/2133$ for $\alpha\neq\beta$ and $A_{\alpha\alpha}=-221/2133$ ($\alpha=e,\mu,\tau$). 

The entropy density, $s$, and the expansion rate of the Universe, 
$H(T)$, are given by:
%%%%%%%%%%%%%%%%%%%%%%%%%%%
\begin{eqnarray}
s  \;= \; \frac{g_{* }2\pi^{2} T^{3}}{45}\,,\quad H(T)\;\simeq\;\frac{1.66\,\sqrt{g_{*}}\,T^{2}}{m_{Pl}}\,.\label{entropy}
\end{eqnarray}
%%%%%%%%%%%%%%%%%%%%%%%%
%
where  $g_*=228.75$  \cite{davidsonetal} and
$m_{Pl}\simeq 1.22\times 10^{19}$ GeV is the Planck mass.
%%%%%%%%%%%%%%%%%%%%%%%%%%%%%%%%%
\begin{figure}
\begin{center}
\includegraphics[width=11cm,height=7.5cm]{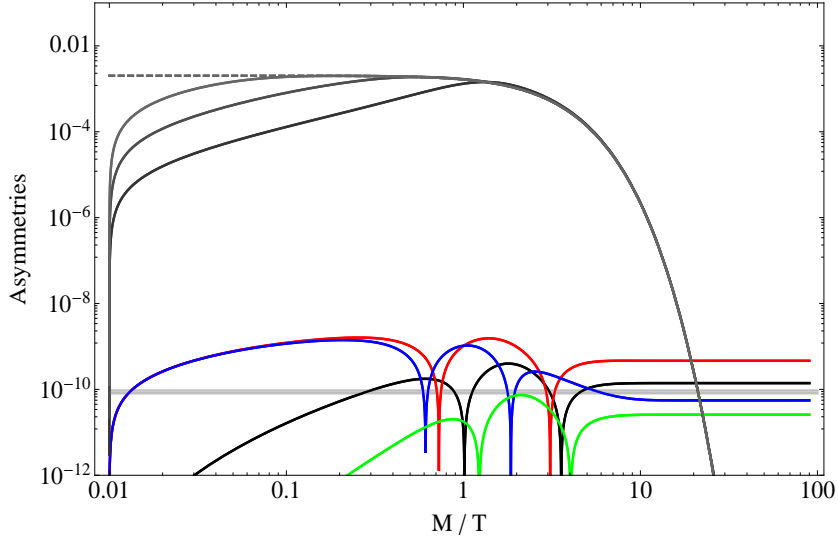} 
\caption{Solution of the  Boltzmann equations  (\ref{YNNBE})-(\ref{YDeltaell}) for $\tan\beta=10$ and $\delta=\pi/4$. See the text for details.
\label{fig1}}
\end{center}
\end{figure}
%%%%%%%%%%%%%%%%%%%%%%%%%%%%%
%

In the case in which the soft SUSY breaking terms are negligible, 
the thermal rates in (\ref{YNNBE})-(\ref{YDeltaell}) satisfy the 
conditions \cite{davidsonetal}: $\gamma_{X}^{i}=\gamma_{X}^{\tilde{i}}$ and 
$\gamma_{X}^{i,\ell}=\gamma_{X}^{i,\tilde{\ell}}=\gamma_{X}^{\tilde{i},\ell}=\gamma_{X}^{\tilde{i},\tilde{\ell}}$. 
As a good approximation, supersymmetric leptogenesis proceeds as a manifest  generalization 
of the standard leptogenesis scenario of the type I see-saw extension of the SM.  Indeed, new effects due to different supersymmetric  equilibration mechanisms between particle and sparticle number
densities provide typically only relatively small corrections \cite{Fong:2010qh}, which can be safely neglected for the purposes of the present study.

The dominant contribution to the production and damping of 
the lepton asymmetries is generally provided by decays and 
inverse decays of $N^\prime_{i}$ \cite{Giudice:2003jh}, whose 
thermal averaged rates are
%%%%%%%%%%%%%%%%%%%%%%%%%%%%%%%%%%%%%%%%%%%%%%%
\begin{equation}\label{gammaD}
 \gamma_{D}^{i}  \; \simeq  \; \frac{M^{3}}{\pi^{2} z} \,\mathcal{K}_{1}(z)\, \Gamma'_{i}\,,\quad\quad \gamma_{D}^{i,\,\ell}\;=\;\gamma_{D}^{i}\,\frac{|(Y'_{\nu})_{\ell i}|^{2}}{({Y'_{\nu}}^{\dagger}Y'_{\nu})_{ii}}\,,         
\end{equation}
%%%%%%%%%%%%%%%%%%%%%%%%%%%%%%%%%%%%%%
%
where $\mathcal{K}_{1}(z)$ is a modified Bessel function of the second kind.

  We neglect in  (\ref{YNNBE})-(\ref{YDeltaell}), for simplicity, 
thermal corrections to the CP asymmetries and the 
decay/scattering rates \cite{Giudice:2003jh}. We do not include either 
the $\Delta L=2$ washout of the flavour lepton asymmetries in 
the Boltzmann equations listed above because they are subdominant 
at the temperatures at which the 3-flavoured leptogenesis 
takes place.~\footnote{As is well known, $\Delta L=2$ scatterings 
mediated by $N^\prime_{i}$ ($\tilde{N^\prime}_{i}$) 
can be safely neglected if 
$\Gamma^\prime_i/H(T)\ll 10\times M_i/ \left(10^{14}\;{\rm GeV}\right)$ 
\cite{davidsonetal}. }

 The final baryon number density 
(normalized to the entropy density of the Universe) is:
%%%%%%%%%%%%%%%%%%%%%%%%%%%%%%%%%%%%%%%%%%
\begin{equation}
Y_{B}\;=\;\frac{10}{31}\,\left( \hat{Y}_{\Delta_{e}} + 
\hat{Y}_{\Delta_{\mu}}+\hat{Y}_{\Delta_{\tau}} \right)\,.
\label{YB1}
\end{equation}
%%%%%%%%%%%%%%%%%%%%%%%%%%%%%%%%%%%
%
In order to have successful leptogenesis, the CP asymmetries 
$\epsilon_{i}^{\ell}$ ($\ell=e,\mu$) should be 
sufficiently large and should have the correct sign. 
According to eq.~(\ref{eif1}), the sign of  
$\epsilon_{i}^{e}=-\epsilon_{i}^{\mu}$ and, consequently,
of $Y_{B}$, is fixed by the value of the rephasing invariant 
associated to the Dirac phase $\delta$, $J_{CP}$. Numerically, from 
(\ref{eif1}), we get for $\tan^2\beta \gg 1$:
%%%%%%%%%%%%%%%%%%%%%%%%%%%%%%
\begin{equation}
	\epsilon_{1}^{e}\;\simeq\;-2.3\times 10^{-6}\,J_{CP}\,(\tan\beta)^{2}\,,\quad \epsilon_{2}^{e}\;\simeq\;1.3\times 10^{-6}\,J_{CP}\,(\tan\beta)^{2}\,,\quad \epsilon_{3}^{e}\;\simeq\;2.1\times 10^{-7}\,J_{CP}\,(\tan\beta)^{2}\,,
\label{epsilonei}
\end{equation}
%%%%%%%%%%%%%%%%%%%%%%%%%%%%
%
where we have used eq. (\ref{mj}) and 
$y_\tau^2 \simeq 10^{-4}\tan^2\beta$. 
Taking, more explicitly, $\tan\beta=10$, one easily obtains:
%%%%%%%%%%%%%%%%%%%%%%%%%%%
\begin{equation}
	\epsilon_{1}^{e}\;\simeq\;-{\rm sgn}(\sin\delta)\,1.4\times 10^{-6}\,,\quad \epsilon_{2}^{e}\;\simeq\;{\rm sgn}(\sin\delta)\,7.0\times 10^{-7}\,\quad
	 \epsilon_{3}^{e}\;\simeq\;{\rm sgn}(\sin\delta)\,1.3\times 10^{-7}\,,
\end{equation}
%%%%%%%%%%%%%%%%%%%%%%%%%%%
%
which, in general, is  the right order of magnitude 
of the CP asymmetry in order to have a successful leptogenesis. 
Notice that ${\rm sgn}(\sin\delta)$ is equal either to $ (-1)$ or to 
$(+1)$, depending on whether the Dirac phase $\delta \cong 5\pi/4$ or 
$\delta \cong \pi/4$, which are the two approximate values 
$\delta$ can have in the model considered
(see eqs.~(\ref{varphi}) and (\ref{varphi2})). 

  Taking into account eq. (\ref{epsilonei}), 
expression (\ref{YB1}) can be recast in the form:
%%%%%%%%%%%%%%%%%%%%%%%%%%%%%%%%
\begin{equation}
Y_{B}\;\approx\; J_{CP}\,(\tan\beta)^{2}\,\epsilon\,\eta_{B}\,
Y_{N^\prime}^{\rm eq}\left( z\ll1\right)\,,
\label{YB2}
\end{equation}
%%%%%%%%%%%%%%%%%%%%%%%%%%%%%%%%%%%%%%%%
%
where $Y_{N^\prime}^{\rm eq} = 45/(\pi^{4} g_{*})\simeq 2\times 10^{-3}$, 
$\epsilon \equiv 10^{-6} $ and $\eta_{B} > 0$ is, by definition, 
the efficiency factor of the asymmetry. 
It follows from eqs. (\ref{JCP}) and (\ref{YB2}) that for 
$\delta = 226.93^{\circ} \simeq 5\pi/4$, the baryon asymmetry
has the wrong sign. Thus, the observed value of the baryon
asymmetry can be obtained in the model considered only for 
$\delta \simeq \pi/4$.

 The the efficiency factor  $\eta_{B}$ in eq. (\ref{YB2}) 
can be computed by solving the full system of Boltzmann equations 
(\ref{YNNBE})-(\ref{YDeltaell}). We note that in the 
model considered the parameter $\eta_{B}$ does not depend 
on the leptogenesis scale $M_{FLG}\sim M$. This can be 
easily understood if one  considers, for simplicity, 
the solution of the Boltzmann equations where only decay and 
inverse decay processes are included: 
as we have already mentioned, this is a good approximation 
in thermal flavoured leptogenesis. 
In this case, from eqs~(\ref{Y'Y'}), 
(\ref{entropy}) and (\ref{gammaD}) one has:
%%%%%%%%%%%%%%%%%%%%%%%%%%%%%%%%%%%%%%%
\begin{equation}
\frac{z\,\gamma_{D}^{i}}{s H(M)}\;
\propto\;\frac{m_{i}\,m_{Pl}}{v_{u}^{2}}\,. 
\end{equation}
%%%%%%%%%%%%%%%%%%%%%%%%%%%%%%%%%%%
%
Therefore, the Boltzmann equations do not explicitly 
depend on the heavy Majorana neutrino mass scale $M$ within the 
indicated approximation. 
We verified numerically that the dependance of $\eta_{B}$ 
and $Y_{B}$ on $M$ is relatively weak also if we take 
into account the scattering processes. 
This implies that, in the class of
SUSY see-saw models of the type considered in this paper, 
the leptogenesis scale $M_{FLG}$ can be lowered sufficiently in order 
to avoid the potential gravitino problem
~\footnote{The Davidson-Ibarra bound \cite{Davidson:2002qv} 
does not apply in the radiative leptogenesis scenario discussed by us.}.

In figure~\ref{fig1}, we report the solution of the full set 
of Boltzmann equations (\ref{YNNBE})-(\ref{YDeltaell}) 
for $\tan\beta=10$ and $\delta=\pi/4$. 
The  red, blue, green and black lines represent 
$|\hat{Y}_{\Delta_{e}}|$, $|\hat{Y}_{\Delta_{\mu}}|$, $|\hat{Y}_{\Delta_{\tau}}|$  and $|Y_{B}|$, respectively. 
The dashed line corresponds to $Y_{N^\prime_{1}}^{\rm eq}$, 
while the other three
black lines are the RH neutrino abundances 
$Y_{N^\prime_{1,2,3}}$. 
The gray horizontal band gives the $3\sigma$ interval of 
experimental values of $Y_{B}$: 
$Y_{B}^{\rm obs}=(8.77\pm 0.21)\times 10^{-11}$ \cite{WMAP}, 
where we have quoted the $1\sigma$ error. 
In this numerical example, we get  the final asymmetries:
%%%%%%%%%%%%%%%%%%%%%%%%%%%%%%%%%%
\begin{equation}
\hat{Y}_{\Delta_{e}}\;\simeq\; 4.7\times 10^{-10}\,, 
\quad\quad \hat{Y}_{\Delta_{\mu}}\;\simeq\;-5.8\times 10^{-11}\,, 
\quad\quad \hat{Y}_{\Delta_{\tau}}\;\simeq\;2.6\times 10^{-11}  \quad\quad\text{and}\quad\quad Y_{B}\;\simeq\; 1.4\times 10^{-10}\,.
\end{equation}
%%%%%%%%%%%%%%%%%%%%%%%%%%%%%%%%%%
%
From eq.~(\ref{YB2}) and the numerical value of $Y_{B}$ thus computed, 
we get an efficiency factor $\eta_{B}\;\simeq\; 0.07$.
Obviously, one can get a value of $Y_{B}$ closer to 
the mean best fit value $\bar{Y}_{B}^{\rm obs} = 8.77\times 10^{-11}$
for a somewhat smaller value of $\tan\beta$.

 We would like to conclude with the following remarks.
As we have shown, the correct sign of the baryon asymmetry in the  
$SU(5)\times T^{\prime}\times Z_{12}\times Z_{12}^{\prime}$ 
model considered \cite{Chen:2007afa,Chen:2009gf} 
can be obtained only in the case of $\delta \cong \pi/4$. 
As has already been discussed in the Introduction, 
for this value of the Dirac phase $\delta$ we have 
$\sin^{2}\theta_{12}\cong 0.37$, 
while the current neutrino oscillation data 
imply at 3$\sigma$ 
$\sin^2\theta_{12} \ltap 0.36$ \cite{Fogli:2011qn}, 
or $\sin^2\theta_{12} \ltap 0.374$ \cite{Maltoni1211}, 
depending on the details of the analysis.   
For $\delta \cong 5\pi/4$, the value of 
$\sin^{2}\theta_{12}\cong 0.299$ predicted by the model 
lies within the $1\sigma$ interval of values
suggested by the data,
but the predicted baryon asymmetry of the Universe 
has the wrong sign
~\footnote{Our result for the sign of the baryon asymmetry 
in the case of $\delta = 226.93^{\circ} \simeq 5\pi/4$
contradicts the claim made in \cite{MCCKMLG0611}.}
(see eq.~(\ref{YB2})).
If $\sin^{2}\theta_{12}\cong 0.37$ would be definitely 
excluded by future data, one would have to modify the 
$SU(5) \times T^{\prime}$ model of flavour
we have considered in the present article.
One possible ``minimal'' modification could be 
to lift the degeneracy in mass of the 
the three heavy Majorana neutrinos (sneutrinos) 
at the scale  $M_{X}$, at which
the flavour symmetry is spontaneously broken.
This could be achieved, e.g., by 
replacing the chiral superfield $S$
in the $SU(5)\times T^{\prime}\times Z_{12}\times Z_{12}^{\prime}$ 
invariant superpotential of \cite{Chen:2009gf} 
with a new chiral supermultiplet $\chi$, 
which is a Standard Model singlet and is charged only under 
the discrete group $Z_{12}^{\prime}$, with charge $\omega^{2}$. 
The model, therefore, has the same gauge and 
flavour symmetry groups and the same number of fields
as the one discussed in  \cite{Chen:2009gf}. 
In this new scenario, the flavour structure
of the superpotential naturally 
generates a Majorana mass matrix (term) for 
the heavy RH neutrinos at the scale $M_{X}$. The latter 
is still diagonalised by the tri-bimaximal
mixing matrix $U_{TBM}$, but has non-degenerate 
eigenvalues. The low energy phenomenology, as well as the 
generation of the baryon asymmetry of this  
class of models is therefore worthwhile investigating,
but such an investigation lies outside the scope of the 
present work.

\section*{Acknowledgments}
We would like to thank M. Spinrath for sharing with us his 
numerical results on the quark mixing observables in the 
$SU(5)\times T'$ model considered.
S.T.P. acknowledges very useful correspondence with R. Gonzalez 
Felipe regarding the resonant radiative leptogenesis.
This work was supported in part by the INFN program on
``Astroparticle Physics'', by the Italian MIUR program on
``Neutrinos, Dark Matter and  Dark Energy in the Era of LHC'' 
and by the World Premier International Research Center
Initiative (WPI Initiative), MEXT, Japan  (S.T.P.). 
The work of E.M. is supported by
Funda\c{c}\~{a}o para a Ci\^{e}ncia e a
Tecnologia (FCT, Portugal) through the projects
PTDC/FIS/098188/2008,  CERN/FP/116328/2010
and CFTP-FCT Unit 777,
which are partially funded through POCTI (FEDER).

\end{document}